\documentclass[11pt]{article}
\usepackage{moriond,epsfig}
\RequirePackage{xspace}

\def\MyPhoto{\psfig{figure=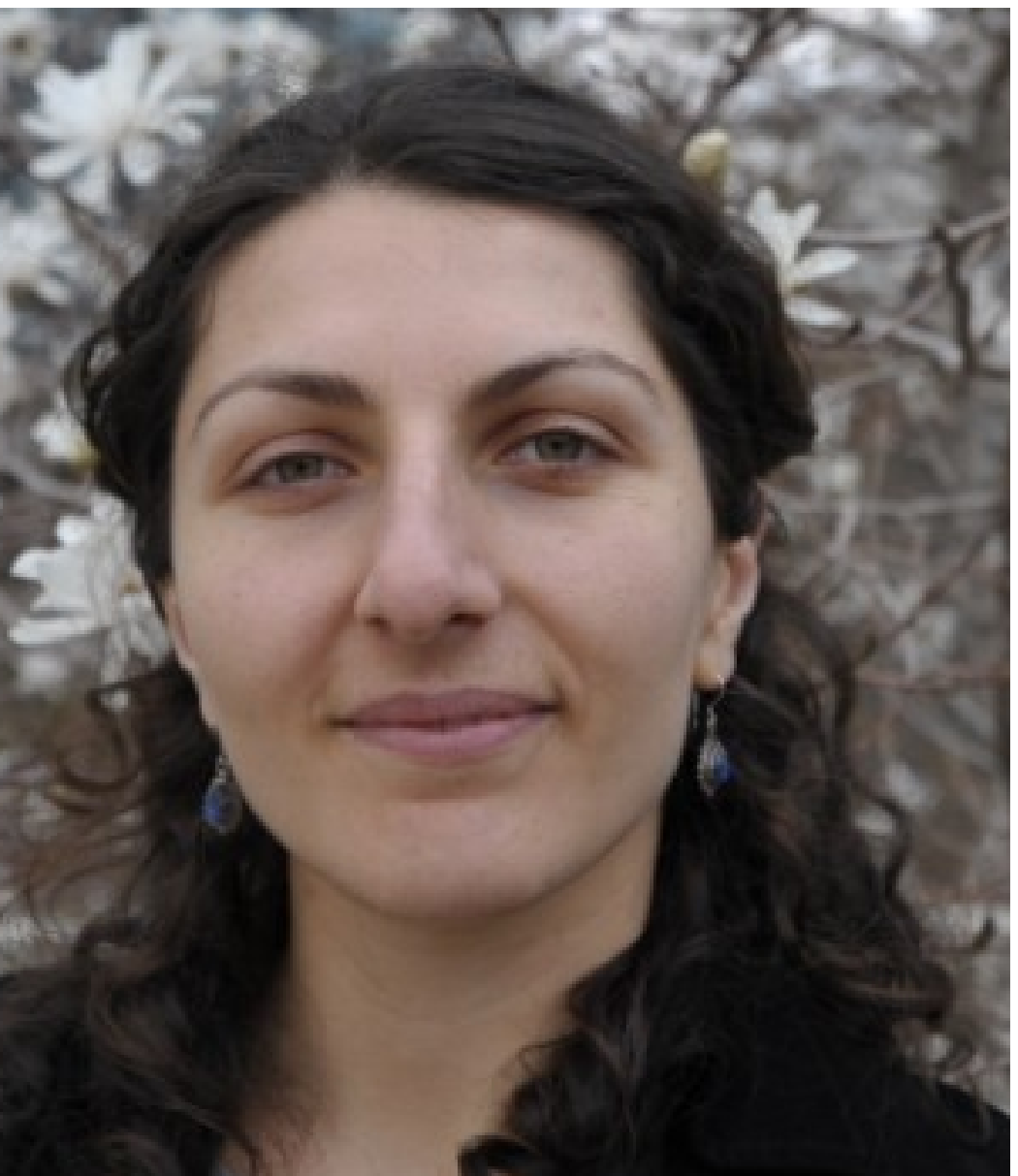,width=35mm}}

\def\Journal#1#2#3#4{{#1} {\bf #2}, #3 (#4)}

\def\PLB{{\em Phys. Lett.}  B}

\def\PRD{{\em Phys. Rev.} D}

\def\ZZ      {\ensuremath{\Z\Z}\xspace}

\def\WZ      {\ensuremath{\W\Z}\xspace}
\def\W      {\ensuremath{W}\xspace}
\def\Z      {\ensuremath{Z}\xspace}

\def\proton      {\ensuremath{p}\xspace}
\def\ellell     {\ensuremath{\ell^+ \ell^-}\xspace}

\def\s     {\ensuremath{s}\xspace}

\def\invpb {\ensuremath{\mbox{\,pb}^{-1}}\xspace}
\def\invfb {\ensuremath{\mbox{\,fb}^{-1}}\xspace}

\begin{document}
\vspace*{4cm}
\title{STUDY OF THE \WZ PRODUCTION AT CMS}

\author{ K. KAADZE }

\address{Department of Physics, Kansas State University, 116 Cardwell Hall,\\
Manhattan, 66506, KS, USA}

\maketitle\abstracts{
In this article we describe the methods of observing \WZ signal 
at CMS experiment at Large Hadron Collider.
To reduce the dependence on Monte Carlo simulation we use data-driven techniques and estimate 
that $5\sigma$ significance of \WZ signal can be reached with less than 350~\invpb at 95~\% C.L. 
for $\sqrt{s}$ = 14 TeV. We also overview several models that would yield anomalous production
of the \WZ final state, sensitivity to which can be achieved with less data.}

\section{Introduction}

The study of multiple gauge boson production at the TeV scale
constitutes a unique opportunity to test the Standard Model (SM) of
electroweak interactions at the highest possible energies.
The production of \WZ events via \s-channel allows to 
probe triple gauge-boson couplings in model-independent manner at the energies 
never attained before and improve our understanding of the gauge theory. 
Any deviation of the strength of these couplings from 
their SM expectations will manifest the new physics. Study of 
\WZ production is also important for new physics searches such as Techicolor, 
Higgsless models, fermiophobic higgs production, $etc.$. 
We briefly describe some of the these models below.

\section{New Phenomena with \WZ Production}
\subsection{Technicolor}

Technicolor (TC) is a strongly interacting gauge theory and one of the models
explaining electroweak symmetry breaking~\cite{TCTheory}. Its recent version has 
slowly-running, ``walking'', couplings which
result in reducing the technicolor scale down to 250 GeV. Thus, a low scale TC 
is more accessible at LHC and the lightest techni-particles $\rho_{TC}$ 
and $a_{TC}$ can have masses below 500 GeV. These particles decay into electroweak bosons, 
for example, $\rho_{TC}/a_{TC} \to \W^{\pm}\Z$.
Another decay channel of $\rho_{TC}/a_{TC}$ is through a production of technipion, 
$\rho_{TC}/a_{TC} \to \W^{\pm}\pi_{TC} \to \W^{\pm} bq$
which has higher branching fraction but experimentally difficult to observe due to severe 
$t\bar{t}$ background at LHC. A detailed study to search for techni-hadrons, $\rho_{TC}$ and $a_{TC}$, 
when \WZ decay into purely leptoninc final state~\cite{CMSTC} shows that 2-10 $\invfb$ of 
integrated luminosity is needed for $5\sigma$ discovery at CMS for different 
mass values of $\rho_{TC}/a_{TC}$.

\subsection{Minimal Higgsless Model}
One of the possible explanations of the electroweak symmetry breaking is extra dimension-based 
minimal Higgsless model~\cite{MHLM}. This model predicts the existence of new 
heavy gauge bosons $\W^{\prime}$ and $Z^{\prime}$ 
which decay into electroweak bosons while decays into fermions are highly suppressed. 
$E.g.$, $W^{\prime}$ decays into \WZ and shares the same final state as the SM \WZ production.

\section{Study of the process $\proton\proton\to \W\Z \to  \ell^{\pm}\nu \ellell$}

The tools and methods developed in order to identify $WZ$ events 
once the data will be available at CMS~\cite{CMS} are described below.
We consider fully leptonic decays of \W and \Z bosons with $3e$, $2e1\mu$, 
$2\mu 1e$, and $3\mu$ charged lepton final state configuration.
These final states have small branching fraction but they have distinct experimental signatures 
which allow to separate signal from copious backgrounds at LHC.
The following processes are dominant backgrounds 
to \WZ production: $\Z+jets$ (the largest), $\Z\gamma$, $\Z\Z$, $W+jets$, and $t\bar{t}+jets$.
The strategy of the analysis is to develop a reliable and efficient lepton identification 
to select \WZ events as well as well-controlled data-driven background estimation 
methods~\cite{CMSWZ} not to rely much on Monte Carlo simulation of the detector.


\subsection{Lepton Identification and Event Selection}
\label{subsec:LepId}
The goal for the lepton identification is to identify electrons and muons
from heavy boson decay and to reject events with misidentified jets and converted photons,
which we refer below as fake leptons.

Electron and muon candidates are identified by matching of the track 
reconstructed by the CMS tracking system to the energy deposition in the 
electromagnetic calorimeter and to the muon track from the muon chambers, respectively. 
Electrons are required to be isolated from activity in the tracker and satisfy
constraint on the shape of energy deposition in the electromagnetic calorimeter. 
Muons must be isolated both in the tracker and the calorimeter and are required to be prompt 
as well to reject the background from semileptonic heavy quark decays.

\WZ events are selected by electron and/or muon triggers and are accepted
if they contain at least three charged leptons, either electrons or muons,
within the detector acceptance having a high transverse momentum. 
The leptons must satisfy selection described above.
A spatial separation between leptons is also required. \Z boson candidate is 
formed from all possible same-flavor, opposite-charge 
lepton pairs, and an event is kept if the mass of the candidate is between 50 and 120 GeV.
The event is rejected if a second independent \Z boson candidate is found.
After \Z boson candidate leptons are identified a lepton from \W boson decay is selected
from the list of remaining leptons that has the highest transverse momentum.
To suppress $\Z+jets$ background when $jet$ is misidentified as an electron from
\W boson decay this electron in addition must be isolated from activity in the calorimeters. 

In order to further reject background from $\Z+jets$ processes we require 
transverse mass of the \W boson candidate to be large
\begin{equation}
\label{eq:wtm}
M_{T}(W) = \sqrt{2\cdot MET \cdot E_{T\ell} (1 - cos\Delta \phi_{MET, \ell})} > 50 \rm~GeV
\end{equation}
Here, MET is the missing transverse energy, $E_{T\ell}$ is the transverse energy of 
the lepton from the \W boson decay, and $\Delta\phi_{MET, \ell}$ is the azimuthal 
separation between the MET and the lepton from \W boson decay.


\subsection{Background Estimation and Signal Extraction}
Backgrounds to \WZ production are separated into three categories:
1) physics background from $\Z\gamma$ and \ZZ production which are estimated 
from Monte Carlo simulation, 2) processes without a genuine \Z boson 
from $t\bar{t}+jets$ and $\W + jets$ production which are ~6\% of the 
\WZ signal and are also estimated from Monte Carlo, and 3) processes with a 
genuine \Z boson from $\Z+jets$ production which is the major background due
to jet being misidentified as a lepton from the \W boson decay.
So-called ``matrix method''~\cite{MM} is used to estimate the contribution
of these processes.

We use data to obtain two sets of samples:
the first sample where the electron from the \W boson decay satisfies track matching, 
energy deposition constraint, and isolation in the tracker, while the muon from 
the \W boson decay satisfies track matching between the tracker and muon spectrometer
is referred to as a ``loose'' sample;
the second sample where the electron from the \W boson decay must satisfy additional
isolation in the calorimeters, while the muon -- the full selection
for muon identification, described in Section~\ref{subsec:LepId}, is referred to as a ``tight'' sample.
Number of events in the former sample, $N_{loose}$, 
consists of events with real isolated leptons $N_{\ell}$ and 
events with fake leptons $N_{j}$:

\begin{equation}
\label{eq:matrixEq1}
N_{loose} = N_\ell + N_j.
\end{equation}

Number of events in tight sample is given by

\begin{equation}
\label{eq:matrixEq2}
N_{tight} = \epsilon_{tight} N_\ell + p_{fake} N_j,
\end{equation}
where $\epsilon_{tight}$ is an efficiency of ``tight''
criteria with respect to ``loose'' requirements for true isolated
leptons and $p_{fake}$ is the same for misidentified jets.

We estimate $\epsilon_{tight}$ from ``tag-and-probe'' method~\cite{TP}
using $\Z \to e^+e^-$ or $\Z \to \mu^+\mu^-$ events. We obtain an efficiency
$\epsilon_{tight}^e=0.98 \pm 0.01$ for electrons and
$\epsilon_{tight}^\mu= 0.98 \pm 0.01 $ for muons.
To determine $p_{fake}$ we use $W+jets$ data sample which has the 
same jet composition as $Z+jets$. 
We obtain $p_{fake} = 0.32 \pm 0.04$
and $p_{fake}^\mu = 0.08 \pm 0.01$ for electrons and muons, respectively.

Plugging the values of $\epsilon_{tight}$ and $p_{fake}$ in Eqs.~\ref{eq:matrixEq1}
and \ref{eq:matrixEq2} we obtain the number of background events $N_{j}$ from $\Z+jet$ 
processes. The results of this method is shown in
Table~\ref{tab:yieldsEstimate}. The results agree well with the expected 
background events from Monte Carlo truth information.

\begin{table}[h]
\caption{Expected number of events for an integrated luminosity of 300 \invpb for the signal
and estimated background for 81 GeV $< M_Z < $ 101 GeV using data-driven methods. Uncertainty is
systematic associated with the background subtraction method only.}
\begin{center}
\vspace{0.4cm}
\begin{tabular}{|lcccc|} \hline 
 & 3e &2e1$\mu$ &2$\mu$1e &3$\mu$\\ \hline
$N$ - ZZ -Z$\gamma$  - W+jets - $t\bar{t}$      &       11.1 $\pm$ 1.3           &8.2 $\pm$ 0.9   &12.1 $\pm$ 1.2   &10.5$\pm$0.8\\ \hline
$N^{genuine~Z}$ (matrix method) &3.2 $\pm$ 1.7           & 0.6 $\pm$ 0.8           &4.6 $\pm$ 2.0   &0.6 $\pm$ 0.9\\ \hline
$N^{\WZ}$                       &7.9 $\pm$ 2.1           & 7.6 $\pm$ 1.2           &7.5 $\pm$ 2.3   &10.0 $\pm$ 1.2\\ \hline
\WZ from MC &7.9&8.1& 9.0 &10.1\\ \hline
\end{tabular}
\label{tab:yieldsEstimate}
\end{center}
\end{table}


\subsection{Signal Significance}

\begin{figure}[tb]
  \begin{center}
  \scalebox{0.4}{\includegraphics{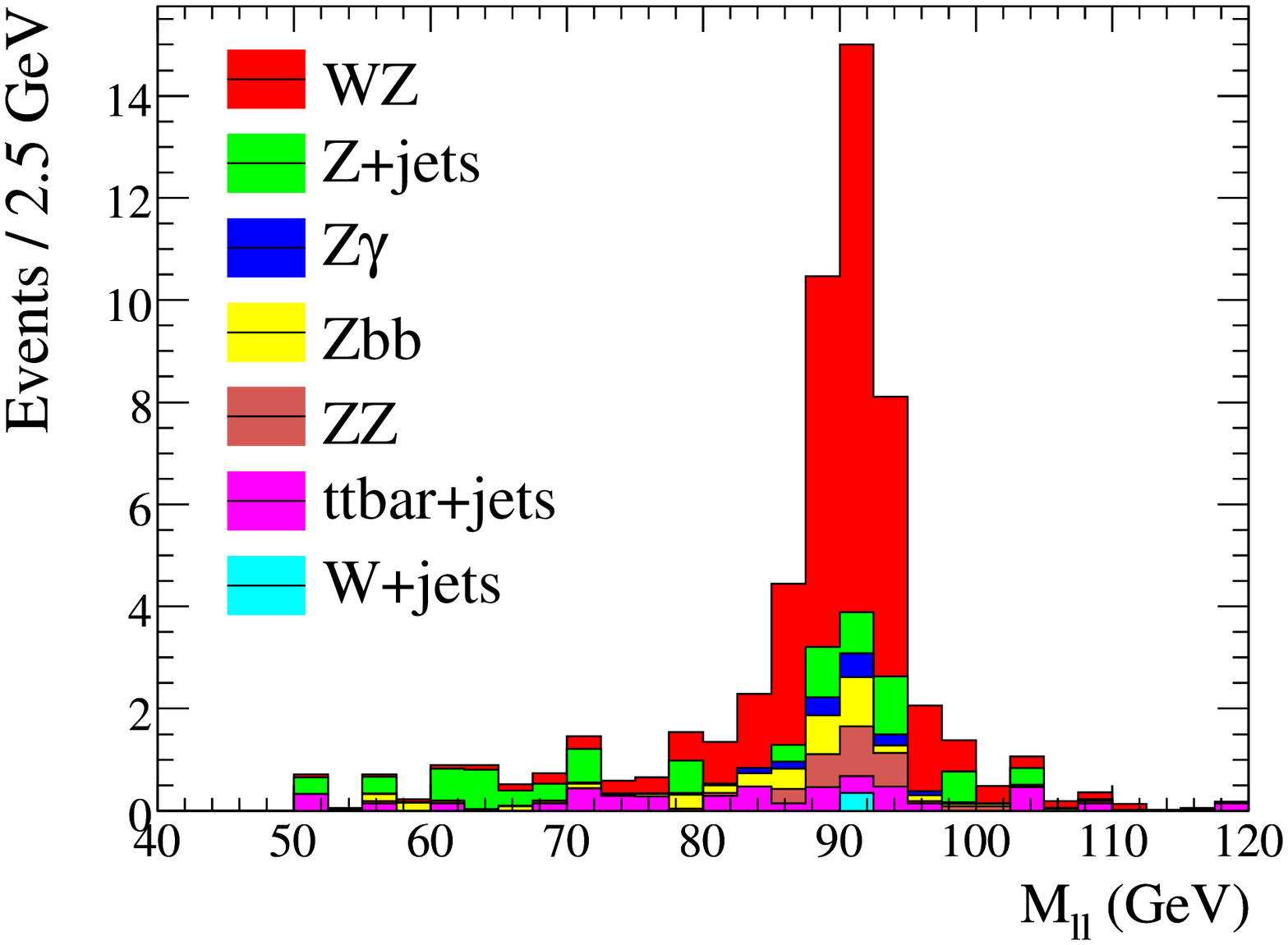}\includegraphics{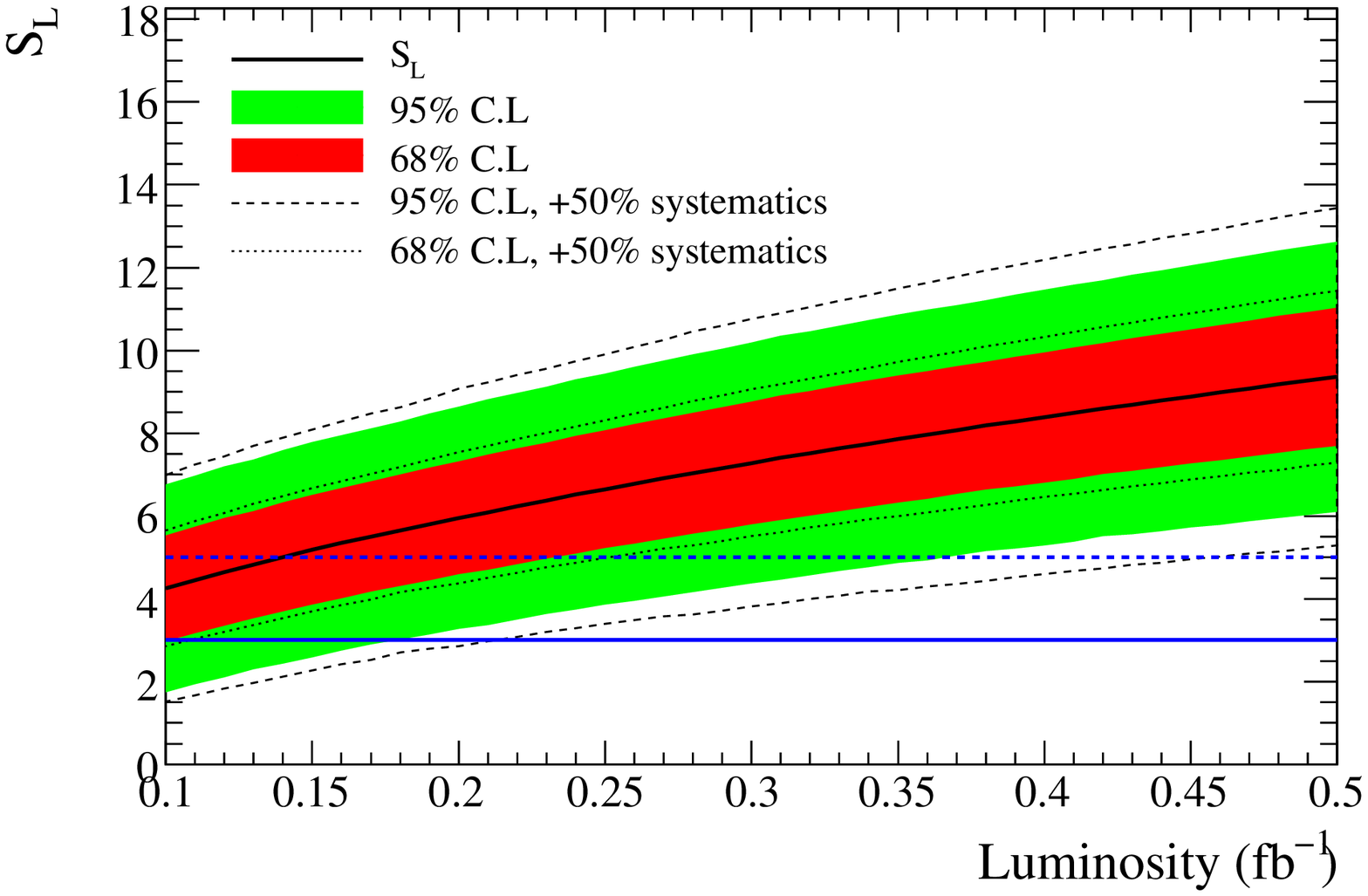}}
  \caption{\Z boson candidate invariant mass for all four channels combined, normalized to
    integrated luminosity of 300~\invpb (left), Expected signal significance for \WZ production
    as a function of integrated luminosity. We use a frequentist approach to estimate variation of expected
    signal and background events. The corresponding 68\% and 95\% C.L. regions are displayed as red and green bands, respectively (right).
  }
  \label{fig:ZMass}
  \end{center}
\end{figure}

The distribution of the \Z boson candidate invariant mass for all four
channels combined after applying the final selection is shown in Fig.~\ref{fig:ZMass}(left).

We estimate the expected signal significance $S_L$ of \WZ production as a
function of integrated luminosity for all four categories
combined using a frequentist approach. The final selection criteria and requirement
of the \Z boson invariant mass to be within 10 GeV from the nominal
\Z boson mass is applied taking into account full systematic and statistical
uncertainties. Result is shown in Fig.~\ref{fig:ZMass} (right). 


\section{Conclusion}

We have studied the methods to establish $\proton\proton \to \W\Z \to \ell^{\pm}\nu\ellell (\ell=e,\mu)$
signal in early data taking at CMS experiment and obtain
the sensitivity for observing the production 
at 95\% C.L. as a function of integrated luminosity. 
It is shown that with less than 500~\invpb of integrated luminosity we 
can achieve $5\sigma$ significance of the signal. We plan to extend this study for 
measurement of the $\W\W\Z$ coupling and search for resonant \WZ production in the future.


\section*{References}
\begin{thebibliography}{99}
\bibitem{TCTheory}K. Lane, S. Mrenna, \Journal{\PRD}{67}{115011}{2003}.

\bibitem{CMSTC}T. Bose, ``A Search for Technicolor in the tri-lepton final state'', CMS CR-2008/004.

\bibitem{MHLM}A. Belyaev {\it et al.}, ``Collider Phenomenology of Higgsless models'', arXiv:0711.1919v.

\bibitem{CMS}CMS Collaboration, ``The CMS experiment at the CERN LHC'', JINST 3:S08004,2008.

\bibitem{CMSWZ}CMS Collaboration, ``Study of the process $\proton\proton\to \W\Z \to  \ell^{\pm}\nu \ellell$ at CMS'',CMS PAS EWK-08-003 (2008).

\bibitem{MM}D0 Collaboration, V.M. Abazov {\it et al.}, \Journal{\PLB}{653}{378-386}{2007}.

\bibitem{TP}CMS Collaboration, ``Measuring Electron Efficiencies at CMS with Early Data'', 
CMS PAS EGM-07-001 (2007).

\end{thebibliography}
\end{document}